\documentstyle [12pt,a4]{article}
\newcommand{\be}{\begin{equation}}
\newcommand{\ee}{\end{equation}}
\newcommand{\bea}{\begin{eqnarray}}
\newcommand{\ena}{\end{eqnarray}}

\newcommand{\sect}[1]{\setcounter{equation}{0}\section{#1}}

\def\Tr   {{\rm Tr}}

\def\Re   {{\rm Re}}
\def\Im   {{\rm Im}}
\def\Ks   {\rlap/K}

\def\dk #1 {{d^{#1}k \over (2\pi)^{#1} 2\omega}}
\def\tp   { \theta_c(x_0-y_0) }
\def\tm   { \theta_c(y_0-x_0) }

\def\(    {\left( }    \def\)   {\right) }
\def\[    {\left[}    \def\]   {\right] }
\def\txt #1 {\qquad {\rm #1} \qquad}
\def\lsim{\; \raise0.3ex\hbox{$<$\kern-0.75em\raise-1.1ex\hbox{$\sim$}}\; }
\def\gsim{\; \raise0.3ex\hbox{$>$\kern-0.75em\raise-1.1ex\hbox{$\sim$}}\; }
\def\@versim#1#2{\lower0.2ex\vbox{\baselineskip\z@skip\lineskip\z@skip
  \lineskiplimit\z@\ialign{$\m@th#1\hfil##\hfil$\crcr#2\crcr\sim\crcr}}}

\begin{document}
\begin{titlepage}
\rightline{CERN-TH.6942/93}

\vskip 3cm

\begin{center}

\bigskip

{\LARGE \bf Introduction to Thermal Field Theory}\\
\bigskip
{\large T. Altherr\footnote{On leave of absence from  LAPP, BP110, F-74941
Annecy-le-Vieux Cedex, France.}}\\
{\em Theory Division, CERN, CH-1211 Geneva 23, Switzerland}\\
\bigskip
\bigskip

{\bf Abstract}
\end{center}
This is intended to be a very basic introduction and short (and of
course incomplete) overview of thermal
field theory. In the first part, I introduce the thermal propagator at
a very simple level and give the Feynman rules using the time-path contour
method. In the second part, I give examples of these rules in scalar
theory and discuss the origin of the thermal mass and other important
effects as infrared divergences and phase transitions. In the third part,
I outline the resummation program of Braaten and Pisarski.
\vfill

\leftline{CERN-TH.6942/93}
\leftline{July 93}
\end{titlepage}

\setcounter{footnote}{0}
\sect{Introduction}

The goal of Thermal Field Theory (TFT) is to describe a large ensemble of
multi-interacting particles (including possible non-Abelian gauge
interactions) in a thermodynamical environment \cite{LvW,Kap}.
The main activity of physicists in that field deals with the description of
the canonical or grand-canonical ensembles, that is of systems at fixed
temperature and chemical potential \cite{Tol}. However, extensions to
non-equilibrium systems also exist [4-7]
 (as well as a description in the micro-canonical ensemble \cite{Wel1}),
but I will not discuss them here.

The TFT approach is different from older and more familiar theories such as
the kinetic theory \cite{Ich,LP} or many-body theory \cite{FW} in the sense
that the TFT is endowed with the following advantages:
i) it uses a path integral approach, ii)
it can treat non-Abelian gauge interactions as QCD and
iii) it is Lorentz-covariant.

Although the TFT is now almost 40 years old, its usefulness has been
particularly acknowledged in the middle of the 70s, for the study of phase
transitions in quantum field theories. Lately,
most of the developments of the TFT \cite{LeB,TFT1} have been
motivated by the study of QCD at finite temperature, as the creation of the
quark-gluon plasma in the laboratory becomes feasible.

Within the TFT, two classes of formalisms can be distinguished: one class
is based
upon a complex-time contour \cite{Kel,Nie,Sch,Mil,KSW} and the other on
$C^*$ algebra \cite{Haa}. I will only describe the former.

The main idea of the TFT is to use the path-integral approach of the usual
vacuum field theory and describe the temperature that appears in the
Boltzmann factor $\exp(-\beta H)$ using the time-evolution operator
with complex time variables $t=-i\beta=-i/T$
(everywhere conventional units $\hbar=c=k=1$ are used).
\bigskip

Historically \cite{Eza}, Matsubara \cite{Mat} was the first to build a TFT
by incorporating a purely imaginary time variable into the evolution operator.
His name is associated with the discrete energy frequencies, the so-called
Matsubara frequencies onto which one has to sum over in the Imaginary-Time
Formalism (ITF). Then came some important contributions from Schwinger
\cite{Sch}, Mills \cite{Mil} and Keldysh \cite{Kel} who developed a
formalism based upon the choice of a contour in the complex plane. This is
called the Real-Time Formalism (RTF). The latest developments include a
functional formulation of the theory \cite{KSW}.
Independently, Umezawa and coworkers
\cite{Ume,Oji} have taken a different approach, based on $C^*$ algebra, called
Thermo-Field Dynamics (TFD), which turns out to give essentially the same
results.

\bigskip
The main applications of the TFT in high energy physics can be sorted in three
classes:

i) {\it Cosmology:} Obviously, if one is interested in the studies of hot
plasmas, the early universe is a very good example. At any time before the
recombination, the expansion length of the universe is much smaller
than the mean free path between the particles, so that one can really speak of
a
perfectly thermalized plasma (at least under certain conditions \cite{EE}).
This is also clearly demonstrated by COBE's
results on the cosmic microwave background radiation, which shows
a purely uniform black-body spectrum, up to fluctuations $\delta T/T \sim
10^{-5}$ \cite{COBE}. The temperature of the
primordial plasma ranges from the Planck mass down to a few eV \cite{Wei}.

In fact, the first and major success of the TFT concerns the
symmetry restoration in spontaneously broken gauge theories at finite
temperature [25-27]. This is also at the basis of the inflationary
scenario \cite{Oli}.
This topic has recently been revived in the case of the electroweak
theory, where the order of the phase transition is crucial for baryogenesis
\cite{Lei}.

There exist other types of study in the early universe, as for instance
the calculation of reaction rates taking place in the hot plasma
in order to determine the abundances of some species \cite{BPPS}.
One should also mention an excellent work by Rebhan on thermal corrections
to quantum gravity \cite{Reb}.

\medskip
ii) {\it Astrophysics:} Cores of neutron stars, supernovae, red giants and
white dwarfs are composed of extremely dense plasmas ($\rho =10^6$-$
10^{15}$g/cm$^3$). The temperature can also be very high during the collapse
of a supernova (a few tens of MeV). There does not exist many applications
of the TFT in such systems, but recent studies on neutrinos and axions emission
rates in these stars have proved to be quite useful [32-38].

\medskip
iii) {\it Heavy-Ion collisions:} This concerns the formation of the
quark-gluon plasma (QGP) in the laboratory \cite{CGS}. Lattice calculations
clearly show that QCD has a phase transition at $T_c =$ 100-200 MeV, above
which hadronic matter is deconfined \cite{Sat}. The TFT is an obvious and
unavoidable tool in the study of the QGP and a lot of work has already been
done in the calculation of signals from such matter [41-43].

\medskip
This is almost all concerning particle physics. Of course, there are other
fields of application as for instance condensed matter, for which TFD was
in fact developed.

To be honest, the plasmon decay into $\nu\bar\nu$ pairs \cite{Bra1}, which is
the dominant cooling mechanism of red giants and white dwarfs, is so far
the only experimental verification (although indirect) of the TFT in
high-energy physics. Therefore, the support of such theories proceeds more
by analogy with what is known from the classical case. It is hoped that
each field cited above will bring, in the near future, observable
verifications to these theories.

\sect{Thermal propagator}

In this section, I will derive the thermal propagator in the simplest
possible model, that is a free scalar field theory. All the relevant features
of the TFT show up already at this simple level, so that extensions to more
sophisticated theories will be straightforward.

The starting point of the TFT is the new definition of an observable $A$ for a
system in contact with a heat bath at fixed temperature $T$ (in the canonical
ensemble):
\be
   \langle A \rangle_\beta = Z^{-1} \Tr \[ e^{-\beta H} A \]
,\ee
where $Z$ is the partition function. The Boltzmann factor weighs the
occupation number of the states that are accessible to the system (here the
trace is performed onto discrete states, but it can of course be continuous).
The next step is to realize that the Boltzmann factor can be described by a
rotation of the operator in the Heisenberg representation
\be
    e^{\beta H} A(t) e^{-\beta H} = A(t-i\beta)
,\ee
where $t$ is now a complex time variable. For analytic reasons, only some
complex time values are allowed. For instance, by evaluating the double
product
\bea
   \Tr\ e^{-\beta H} A(t) B(t')
   &=& \sum_m e^{-\beta E_m} \langle m| A(t) B(t') |m\rangle  \nonumber\\
   &=& \sum_{n,m} e^{-iE_n(t-t')} e^{iE_m(t-t'+i\beta)}
      \langle m|A(0)|n\rangle\langle n|B(0)|m\rangle
,\ena
it appears that the trace operation is meaninful only for time arguments
in the range
\be
        -\beta < \Im  (t-t') < 0
.\ee
Consider now the 2-point Green function in a scalar theory. By
definition
\bea
     G(X-Y) &=& \langle T_c(\phi(X) \phi(Y)) \rangle_\beta \nonumber\\
            &=& \tp G_+(X-Y) + \tm G_-(X-Y)
,\ena
where a complex time ordering has been defined ($\tp = 1$ if $y_0$
precedes $x_0$ along the contour).
The domain of analyticity of the 2-point Green function is given by
\bea
  -\beta \le \Im (x_0-y_0) \le 0     &{\rm when}&  \tp=1    \nonumber\\
       0 \le \Im (x_0-y_0) \le \beta &{\rm when}&  \tm=1
.\ena
This means that the contour in the complex plane must
not go up but can only go down.

An important property, the Kubo-Martin-Schwinger (KMS) relation, can also be
derived:
\be
      G_+(x_0-(y_0+i\beta),\vec x-\vec y) = G_-(X-Y)
,\ee
which follows from the trace invariance by circular permutation.

\bigskip
We are now in a position to write down the 2-point Green function for a
free scalar field in terms of operator kernels. Consider a superposition of
creation and annihilation operators in the plane-wave approximation
\be
\phi(X) = \int {d^3k \over{(2\pi)^{3\over2}}} {1\over{\sqrt{2\omega}}}
              [ a^{\dag}(K) e^{iK\cdot X} + a(K) e^{-iK\cdot X} ]
,\ee
where $\omega = k_0 = \sqrt{|\vec k|^2 + m^2}$.
One finds for the thermal propagator
\bea
G(X-Y) &=& \int{d^3k\over (2\pi)^{3\over2} \sqrt{2\omega_k}}
           \int{d^3p\over (2\pi)^{3\over2} \sqrt{2\omega_p}}   \\
\times &\bigg\{ &\tp \[
   e^{-iK\cdot X+iP\cdot Y} \langle a(K)a^{\dag}(P) \rangle_\beta
 + e^{ iK\cdot X-iP\cdot Y} \langle a^{\dag}(P)a(K) \rangle_\beta \]
\nonumber\\
 &+ &\tm \[
   e^{ iK\cdot X-iP\cdot Y} \langle a(K)a^{\dag}(P) \rangle_\beta
 + e^{-iK\cdot X+iP\cdot Y} \langle a^{\dag}(K)a(P) \rangle_\beta \]
                        \bigg\}\nonumber
.\ena
There remains to calculate the thermal kernels.
Taking for Hamiltonian
\be
   H = \int {d^3 k\over (2\pi)^3} {\omega\over 2}
                 : a(K)a^{\dag}(K) + a^{\dag}(K)a(K) :
,\ee
one can start from the thermal average of the commutation relation
\be
  \langle a(P)a^{\dag}(K)\rangle_\beta
  - \langle a^{\dag}(K) a(P)\rangle_\beta = \delta^3(p-k)
.\ee
And using the cyclicity of the trace and the commutation relation
$[H,a]=-\omega a$, one arrives at
\bea
\langle a^{\dag}(K)a(P) \rangle_\beta &=& n_B(\omega)\delta^3(p-k)
\nonumber\\
\langle a(P)a^{\dag}(K) \rangle_\beta &=& (1+n_B(\omega))\delta^3(p-k)
,\ena
where I have defined the Bose-Einstein statistical weight
\be
         n_B(\omega) = {1\over e^{\beta\omega} - 1}
.\ee
The propagator can then be rewritten as
\bea
G(X-Y) = \int\dk {3} &\bigg[ & \tp e^{-iK\cdot (X-Y)} + \tm e^{ iK\cdot (X-Y)}
\nonumber\\
                 && + n_B(\omega) \( e^{ iK\cdot (X-Y)} + e^{-iK\cdot (X-Y)} \)
                        \bigg]
.\ena
At this level, one has to specify the contour. As the contour must
start from some initial time $t_0$ and go down to $t_0-i\beta$, the
simplest choice is just a straight line along the imaginary-time axis.
This choice of contour leads to the ITF and is
the oldest and most widely used formalism in the TFT. Parametrizing
$x_0-y_0=-i\tau$, one has
\be
G(\tau,\vec x -\vec y) = \int\dk {3} e^{-i\vec k.(\vec x -\vec y)}
          \[ (1+n_B(\omega))e^{-\omega|\tau|} + n_B(\omega)e^{\omega|\tau|} \]
.\ee
In momentum space, the theory reduces to a 3-dimensional euclidean theory
with an infinite summation over the Matsubara frequencies $\omega_n=i2\pi n T,
n \in ]-\infty,+\infty[$. Explicit calculations show however that is more
convenient to stay in $\tau$-space \cite{Pis1}.

The obvious disadvantage of this contour is to lose completely the
real-time argument. In principle, with this formalism, one is restricted to
calculate static thermodynamical quantities
as the free energy. This has motivated several people to consider a
different contour, which would keep a real-time argument.

The contour of the RTF is shown in fig.~1 and is composed
of four different pieces. This is the simplest choice of contour if one
wants to have real-time arguments. For analytic reasons, it can be shown
that the contributions from $C_3$ and $C_4$ can be neglected
\cite{LvW,Eva1}. One is therefore left with two possibilities for both $x_0$
and $y_0$ to lie either on
$C_1$ or $C_2$. This gives four different propagators, which are usually
written in a matrix. They are labelled as ``11'' propagator,
corresponding to $x_0 \in
C_1$ and $y_0 \in C_1$, and so on. An appropriate choice of the parameter
$\sigma$ is the intermediate value $\sigma=\beta/2$. In this case, one
finds the following form of the complete propagator in momentum space
\be
  G(K) = \( \begin{array}{cc} G^{11}(K) & G^{12}(K) \\
                              G^{21}(K) & G^{22}(K) \end{array} \)
       = U(\beta,K) \(
           \begin{array}{cc}
                         \Delta(K) & 0    \\
                         0         & \Delta^*(K)
           \end{array} \)
           U(\beta,K)
,\ee
where
\be
    U(\beta,K) =
       \( \begin{array}{cc} \cosh{\theta_K} & \sinh{\theta_K} \\
                            \sinh{\theta_K} & \cosh{\theta_K} \end{array} \)
,\ee
and
\be
    \cosh{\theta_K} = \( 1 - e^{-\beta\omega} \) ^{-1/2}  \qquad ,\qquad
    \sinh{\theta_K} = e^{-\beta\omega/2} \( 1 - e^{-\beta\omega} \) ^{-1/2}
.\ee
In (2.16), $\Delta$ is the Feynman propagator at zero temperature
\be
  \Delta(K) = {i\over K^2 - m^2 + i\epsilon}
.\ee
The welcome surprise is that these are just exactly the propagators in TFD.
This is a very strong support of the theory in the sense that TFD and the
RTF do not use at all the same mathematical framework.

In TFD language, the 2-type propagators: the ``12'', ``21' and ``22'',
 are associated to ``ghost'' fields
(not to be mistaken for Fadeev-Popov ghosts!). They
are unphysical since one of the time arguments has
an imaginary component. The only physical propagator is the ``11''
component
\be
G^{11}(K) = \Delta(K) + 2\pi n_B(\omega)\delta(K^2-m^2)
.\ee
This propagator was derived in the early attemps at real-time techniques
\cite{DJ}.
It clearly shows the
finite-temperature  contribution in the $\delta$-piece.
But, in order to have a consistent theory, the other propagators must also
be taken into account.

The Feynman rules in the RTF are the following: diagrams have the
same topology as in the vacuum theory, and the same symmetry factors.
Furthermore, there are
two types of vertices, 1 and 2. Physical legs must always be attached to
type-1 vertices, never to type-2 ones. One must sum over all the
configurations of type-1 and type-2 vertices according to the above rule. Then,
one has to use a $G^{ab}(K)$ propagator between a vertex of type $a$ and a
vertex of type $b$ when $K$ flows from $a$ to $b$. All the rest remains the
same as in the vacuum theory.

\medskip
The Feynman rules shown above are those of a free-field theory.
Implementing interactions is rather straightforward and is done in the
same way as in vacuum, as all the necessary formalism is already set up.
For cubic, $(g/3!) \phi^3$, or quadratic, $(g/4!) \phi^4$, interactions,
the Feynman rule is to attach $-ig$ at a type-1 vertex and $+ig$ at a
type-2 one.
In fact, in order to construct these rules one has to make the
assumption that the
interaction is adiabatically swichted on and off around $t=0$,
which is just the usual assumption in vacuum theory \cite{Abr}.
In a heat bath, this is certainly questionable, as one might think of any
particle in constant
interaction with its thermal surrounding. One can phrase the question
differently by asking: are there asymptotic states in a heat bath?
Indeed, practical calculations show some difficulties \cite{AALBP} and formal
considerations seem to give a negative answer \cite{Lan}.
Still, perturbative calculations can make sense, as I will try to show
in the next sections.

\bigskip
The field theory at finite temperature is renormalizable, provided the
vacuum theory is so. This is intuitively obvious as the thermal corrections
all come with a Boltzmann factor that cuts off any ultraviolet divergence.
At the perturbative level, the problem appears more subtle, but is clearly
not an issue.

\bigskip
The path-integral formulation of the same theory is a straightforward
exercise
\cite{KSW}. The generating functional is given by
\be
  Z[j] = Z[0] \langle T_c \exp\( i\int_C d^4 x j(x)\phi(x) \) \rangle_\beta
,\ee
and the path is such that $x(t-i\beta)=x(t)$. By differentiation with respect
to the sources, one obtains the $n$-point Green functions:
\be
  G^n(x_1,\ldots,x_n) = {1\over Z[0]} {\delta^n Z[j] \over i\delta j(x_1)
                      \ldots i\delta j(x_n)} \bigg|_{j=0}
.\ee
With the contour $C$ shown in fig.~1, the same matrix propagator as in
eq.~(2.16) is easily recovered. For completeness, I also give the gauge
boson ``11'' propagator in Feynman gauge
\be
G^{11}_{\mu\nu}(K) = -g_{\mu\nu} \( \Delta(K) + 2\pi n_B(\omega)\delta(K^2)\)
.\ee
Similar relations hold for the other components (one just has to multiply
the thermal matrix (2.16) by the metric). For Dirac fermions, one has
\be
S^{11}_F(K) = (\gamma_\mu K^\mu + m)
\( \Delta(K) - 2\pi N_F(k_0)\delta(K^2-m^2)\)
,\ee
where the Fermi-Dirac statistical weight has been introduced (in presence of
a chemical potential $\mu$)
\be
N_F(k_0) = {\theta(k_0) \over e^{\beta(k_0-\mu)} + 1}
         + {\theta(-k_0) \over e^{\beta(-k_0+\mu)} + 1}
.\ee
Again, one has to multiply the Dirac structure to the fermionic scalar
matrix in order to have the other components. One should also
notice that the ghost gauge fields, although anticommutant, obey the
Bose-Einstein statistics.

\bigskip
After some initial confusion, there has been a considerable amount of work
in the past few years in comparing the different formalisms. The first
remark is that, by using the RTF with a free parameter $\sigma$, one gets
the TFD for $\sigma=\beta/2$, and the Keldysh, or also called the ``cut''
propagators, for $\sigma=0$ or $\sigma=\beta$. The comparisons
between RTF and ITF are
much more complicated, since the Feynman rules are rather different
\cite{FGN}. However, there is now a general agreement that they give the
same physical answers \cite{RTFITF}. In fact, it is
also possible to calculate
dynamical quantities in the ITF, by analytic continuation to real-time
values $\omega \to k_0+i\epsilon$. Note that the equivalence between the
two formalisms is obtained by taking the analytic continuation
$\omega \to k_0+i\epsilon k_0$.
The question of which formalism is the easiest to use is a difficult one.
Each has its advantages and its weaknesses, and using one or
the other is probably more a matter of taste than of convenience.
However, one should note that when dealing with non-equilibrium systems, a
real-time approach is mandatory. For systems at equilibrium, the best
solution is certainly to use both formalisms, as this provides a useful
check on the calculations.

\sect{Some examples in scalar theory}

\subsection{Tadpole diagram and comparison between ITF and RTF}
The simplest thing to calculate in the TFT is the tadpole diagram in a scalar
theory with quadratic interactions as $g\phi^4$ theory.
Is may also be the most instructive calculation.
The Feyman rules in the ITF give
\be
\Sigma = {g\over 2} T \sum_{n=-\infty}^{+\infty} \int {d^3 k\over (2\pi)^3}
         {1\over \omega_n^2 + {\bf k}^2 + m^2}
,\ee
where $\omega_n=2in\pi T$. In order to perform the summation over $n$,
 the
standard trick is to use a contour integral
\be
T \sum_{n=-\infty}^{+\infty} f(k_0=i\omega_n) = {T\over 2i\pi} \int_C dk_0
f(k_0) {\beta\over 2} \coth{\beta k_0\over 2}
,\ee
provided $f(k_0)$ has no poles on the imaginary axis. The contour can be
deformed in such a way that (see fig.~2)
\be
T \sum_{n=-\infty}^{+\infty} f(k_0=i\omega_n) = {1\over 2i\pi} \[
    \int_{-i\infty}^{+i\infty} dk_0 f(k_0)
  + \int_{+i\infty}^{-i\infty} dk_0 (f(k_0) +f(-k_0)){1\over e^{\beta k_0}-1}
\]
.\ee
This allows us to separate the vacuum part (first term) from the
finite-temperature contribution (second term).
Using the explicit value of $f(k_0)=1/(\omega^2-k_0^2)$ with
$\omega=\sqrt{k^2+m^2}$, one gets for the thermal contribution to the tadpole
diagram
\be
\Sigma_\beta = {g\over 2} \int {d^3 k\over (2\pi)^3}
               {1\over \omega} n_B(\omega)
.\ee
This result can be derived in a much simpler way by using the ``mixed''
propagator \cite{Pis1}
\bea
\Sigma &=& {g\over 2} T \sum_{n=-\infty}^{+\infty} \int {d^3 k\over (2\pi)^3}
         \int_0^\beta d\tau e^{i\omega_n\tau} \Delta(\tau,\omega)  \nonumber\\
       &=& {g\over 2} \int {d^3 k\over (2\pi)^3}
         \int_0^\beta d\tau \delta(\tau) \Delta(\tau,\omega)  \nonumber\\
       &=& {g\over 2} \int {d^3 k\over (2\pi)^3}
         {1\over 2 \omega} (1+2n_B(\omega))
.\ena
The ``1'' piece is again the vacuum contribution, which is quadratically
ultraviolet-divergent while the ``$2n_B$'' piece is the finite-temperature
contribution.

With the RTF, the separation from $T=0$ to $T\ne 0$ is
automatic from the beginning, so that one can write
\be
\Sigma_\beta = {g\over 2} \int {d^4 K\over (2\pi)^4}
               2\pi\delta(K^2-m^2) n_B(\omega)
.\ee
The integration on $k_0$ can be trivially performed with the help of the
$\delta$-function and the same result is obtained as with the ITF.

This one-loop example does not illustrate well the complexity and the relative
advantages of each technique, which are observed at higher orders. One can
just see that the methods are quite different.

\subsection{High-temperature expansion and thermal mass}
With this one result in hand, let me continue the discussion. We have
\be
\Sigma_\beta = {g\over 4\pi^2} \int_m^\infty {k\ d\omega
               \over e^{\beta\omega} - 1}
.\ee
The remaining integral cannot be computed analytically, except in some
limiting cases. Let me discuss the high-temperature limit, which
is of greatest interest. One has
\be
\Sigma_\beta = {g\over 24} T^2 \( 1 - {3\over\pi}{m\over T}
                              - {3\over 2\pi^2}{m^2\over T^2}\ln{m\over T}
                                    +\ldots \)
.\ee
The tadpole diagram is the first-order correction to the free propagator
and therefore the above result is nothing but a mass term (because it is
independent of the momentum). In the limit of zero bare mass, one has
the contribution
\be
m_\beta^2 = {g\over 24} T^2
.\ee
This is an {\it effective} mass and corresponds to the fact that in the
heat bath, the propagation of particles is altered by their continuous
interactions with the medium. This is certainly one of the most important
results of the TFT.

The fact that even massless particles acquire a mass does not affect the
chiral symmetry. The thermal mass is generated radiatively and comes from a
term that is chiral-symmetric if the initial Lagrangian is.

\subsection{Higher orders and infrared problems}
In order to see if the perturbation series is well behaved, it is
necessary to look at higher orders. At two-loop order, two different
topologies contribute (see fig.~3). According to the RTF rules, one has
for the first topology
\be
\Sigma^{(2)}_a(K)=-i{g^2\over 4}\int {d^4 P\over (2\pi)^4} D_{11}^2(P)
                      \int {d^4 K\over (2\pi)^4} (D_{11}^2(K) - D_{12}^2(K))
,\ee
where I have used the fact that the type-2 tadpole is identical to the
type-1 tadpole. It is interesting to note that
\be
D_{11}^2(K) - D_{12}^2(K) = \Delta^2(K) - 2PP{1\over K^2-m^2}
                 2\pi\delta(K^2-m^2)n_B(\omega)
.\ee
Thus, even if there are multiple products of statistical weights at
intermediate stages of the calculation, they disappear when combining the
different type-1 and 2 vertices. Also disappear the ill-defined
$\delta$-products. This is a generic feature of the RTF. In fact, a general
theorem can be stated: there appears at most only one statistical factor
(or a sum of) in a loop integral. This is obvious in the ITF, because it is
the summation over the Matsubara frequencies that gives the statistical
weight.

As usual, the ultra-violet singularities
in eq.~(3.10) can be eliminated by
choosing some suitable counter-terms (in fact the same counter-terms as at
$T=0$). Notice, however, that all topologies must be added in order to
cancel the ultra-violet divergences, with a non-trivial mixing between
$T$-dependent and $T$-independent terms \cite{Alt2}. In the high-temperature
limit, the contribution to the thermal mass becomes
\be
\delta m_\beta^{2\ (2)} = - {g^2 \over 384\pi} T^2 \( {T\over m} + \ldots \)
,\ee
and is infrared-singular when $m\to 0$. This power-like singularity
originates solely from the first topology in fig.~3. Although infrared
singularities are inherent in almost all perturbation theories, whether at zero
or finite temperature, there are good reasons to worry about this one:
i) the above contribution enters as a correction to the
thermal mass, and therefore participates into the solution of the pole of
the propagator, which is supposed to be a physical quantity \cite{KKR};
ii) at zero temperature, famous theorems as the
Kinoshita-Lee-Nauenberg theorem \cite{Kin,LN}
demonstrate that singularities appearing at intermediate stages of the
calculation cancel in the final physical result. However, a crucial
argument of the Lee-Nauenberg version of the theorem is the finiteness of
the mass shift at all orders \cite{LN}, which is just the quantity considered
here;
iii) indeed, this mass shift will enter into physical quantities as it
acts as a kinematical cut-off and will appear, say, in the production rate
of light weakly coupled particles from the heat bath.

So, what can be done with this infrared singularity? If it cannot be
cancelled, then one solution is to use an infrared regulator.
In fact, an obvious infrared regulator is just the thermal mass obtained
at first order, $m_\beta$. In that case, the resulting behaviour is
\be
\delta m_\beta^{2\ (2)} = O(g\sqrt{g} T^2)
.\ee
That is, the result is still perturbatively acceptable as it is smaller than
the
previous correction. Unfortunately, this hope does not survive higher-order
corrections. Indeed, by attaching an arbitrary number of tadpoles to
the first line (see fig.~4), one gets always the same correction
\be
\delta m_\beta^{2\ (N)} = O(g\sqrt{g} T^2)
.\ee
An infinite number of diagrams contribute to the same order in the coupling
constant. This clearly shows the breakdown of perturbation series at finite
temperature.

On the other hand,
 this infinite subset of the most infrared diagrams can be
resummed. It is nothing but a Taylor expansion \cite{Alt2}
\be
\sum_{N=0}^\infty {m_\beta^{2N}\over N!} \( {\partial \over
\partial m^2} \) ^N \int_m^\infty {k\ d\omega\over e^{\beta\omega}-1}
\bigg|_{m=0} = \int_{m_\beta}^\infty {k\ d\omega\over e^{\beta\omega}-1}
.\ee
The right-hand side of this equation is perfectly defined for any value of
the coupling constant $g$. Seen from this perspective, the infrared
problem originates from the bad expansion of $\sqrt{g}$ around zero.

The resummed result is not very different from the first-order correction,
$m_\beta$, which, in the light of the above equation, is not surprising.
This seems to make perturbation theory at finite temperature reliable.
I will come back to this important point in the last section.

\subsection{Phase transitions}
The appearance of a thermal mass has another important consequence.
Consider a bare Lagrangian given by
\be
{\cal L}(\phi) = {1\over 2} (\partial_\mu \phi)^2 + {1\over 2}\mu^2\phi^2
            - {g\over4!} \phi^4
.\ee
This Lagrangian has a {\it negative} square-mass term. The minimum of the
effective potential is degenerate and occurs for non-zero values of the
field. Now, the temperature introduces a {\it positive} square-mass term.
The effective potential at high temperature then looks very different
\be
V(\phi_c) = {g\over 48}T^2\phi_c^2 + {g\over4!} \phi_c^4
,\ee
as it has only one minimum at $\phi_c=0$. Hence, there must be a {\it phase
transition} between the high- and the low-temperature regime. Whether this
phase transition is of first or of second order is another question,
which requires to take into account some subleading terms \cite{Lei}.
Still, this is an important result, which has truly modified our way of
thinking about the early universe. Any symmetry that is spontaneously
broken in our present world, which is a low-temperature system ($T=$ 3 K),
will be eventually restored if going to sufficiently early times. This
is true for instance for the Higgs mechanism, which is responsible for all the
particle masses. Above $T\gsim 1$~TeV, all particles become massless.

\bigskip

This feature is not specific to $g\phi^4$ theory.
With $g\phi^3$ interactions in a 6-dimensional space-time, the theory
becomes completely unstable above a certain critical temperature, which can be
computed exactly \cite{AGP}. Phase transitions do also exist in gauge
theories \cite{Lin1,Lei}.

\sect{The resummation of Braaten and Pisarski}

\subsection{The idea}
We have seen in the previous section that in order to get rid of infrared
problems, one has to resum a certain class of diagrams. However, as shown
above in the case of the thermal mass, it is not always necessary to
perform this resummation if one is not interested in higher-order
corrections. In the following, I will present a set of simple rules that
determine in which case one has to resum infrared diagrams. The method is
due to Braaten and Pisarski \cite{Pis2,BP1} and works also for gauge theories.

Consider $g^2\phi^4$ theory at very high temperature, so high that all
other scales as for instance the bare mass, can be neglected. Still,
possible other
scales can be generated radiatively as $gT$, or ``non-perturbatively'' as
$(\ln{g}) T$. The starting point of the BPR is to make a distinction between
{\it hard} momenta of order $T$ and {\it soft} momenta of order
$gT$. The second step is to realize that the thermal mass is generated by a
loop integral, where the momentum running inside the loop is hard. This is
clearly what gives the $g^2 T^2$ contribution to $m_\beta^2$. Soft momenta
give higher-order contributions as $g^3 T^2$ terms. The hard momentum
contribution to the loop integral is called a {\it hard thermal loop} (HTL).
The third and final step consists in resumming only soft lines. Indeed,
the corrections to the bare propagator $K^2$ being of order $g^2T^2$,
they start to be relevant only when $K\sim gT$, which is the soft scale.
Hard lines do not need to be resummed and one can still use the bare
perturbation series.

With these simple rules, one has an improved perturbation theory.
For an $n$-point Green function, the dominant contribution always comes
from the HTL, which is of order $g^n T^2$. If there is no HTL
inside the loop, the integral must be cut into two pieces,
a first piece where the momentum running inside the loop is hard, for which
one uses bare propagators, and a second piece where one has to use resummed
propagators as the loop-momentum becomes soft. The final result should be
independent of the arbitrary intermediate cut-off of order $\sqrt{g}T$,
which is put by hand to separate the two pieces.

In $g^2\phi^4$ theory, there is only one HTL, which is the tadpole
diagram. Therefore only the 2-point Green function needs to be
resummed. The effective Lagrangian that leads to
the improved perturbation series can be written as \cite{BP2}
\be
{\cal L}_{eff} = {1\over 2} (\partial_\mu \phi)^2 - {1\over 2} m_\beta^2 \phi^2
,\ee
with $m_\beta$ given by eq.~(3.9).

The very nice surprise of the BPR is that these features survive when
considering gauge field theories. In QED, the 2-,3- and 4-point
Green functions have an HTL. The same is true in QCD, where
the $n$-point Green function with only external gluon lines, or with
$n-2$ gluon lines and 2 quark lines, must also be taken into account.
In these cases though, HTLs are not just scalars but complicated analytic
functions of the external momenta, $g^n T^2 f(\omega_i,k_i)$. Another
surprise is also that these functions are the same for QED and QCD (at
least for the same number of bosonic and fermionic external lines).
Finally, there are two extra bonuses: i) the HTLs are gauge-invariant
and ii) they obey simple Ward identities.

The HTLs can be derived from an effective Lagrangian given by \cite{BP2}
\be
{\cal L}_{eff} = {1\over 2} F_{\mu\nu}F^{\mu\nu} + {\bar \psi} \gamma_\mu
                 D^\mu \psi
        + {3\over 2}\omega_0^2 F^{\mu\alpha} \langle {K_\alpha K_\beta
  \over -(K.D)^2} \rangle F^{\beta\mu} + m_\beta^2 {\bar \psi} \langle
  {\gamma_\mu K^\mu\over K.D} \rangle \psi
,\ee
where $\langle\rangle$ refers to an angular average; $K$ is the hard loop
momentum; the two parameters $\omega_0$ and $m_\beta$ are the plasmon
frequency and the thermal fermion mass respectively (of order $gT$).
With this Lagrangian, one has an improved perturbation series defined
in terms of effective propagators and vertices (they are pictured in
fig.~5 in the case of QED).

In the next section, I shall present and discuss the HTLs for the bosonic and
fermionic self-energies.

\subsection{The bosonic hard thermal self-energy}
An important consequence of the fact that the plasma constitutes a
privileged rest frame is that the polarization operator can be decomposed
into two different propagating modes, transverse and longitudinal (contrary
to the vacuum case, both are physical):
\be
\Pi_{\mu\nu}(K) = P_{\mu\nu} \Pi_T(K) + Q_{\mu\nu} \Pi_L(K)
.\ee
Expressions for the projectors can be found in \cite{Wel2}.
The HTL is obtained by taking the high-temperature limit and by considering
a soft external momentum, $\omega, k \ll T$. After renormalization of the
ultra-violet zero temperature piece, the thermal corrections for each mode
are
\bea
\Pi_T(K) &=& {3\over 2}\omega_0^2 \( {\omega^2\over k^2}
   + \( 1 -{\omega^2\over k^2}\) {\omega\over 2k}\ln{\omega+k\over \omega-k} \)
\nonumber\\
\Pi_L(K) &=& 3\omega_0^2 \( 1 - {\omega^2\over k^2} \)
              \( 1 - {\omega\over 2k}\ln{\omega+k\over \omega-k} \)
,\ena
where $\omega_0$ is given by
\be
    \omega_0^2 = \( {N_f\over 2} + N \) {g^2 T^2 \over 9}
,\ee
for an $SU(N)$ gauge theory with $N_f$ fermions. For QED, one has to
substitute $(N_f/2 + N ) g^2 \to e^2$. Identical expressions are obtained
in the ultra-degenerate limit, $\mu\gg T$, except for $\omega_0$, which
scales with $\mu$ instead of $T$. When the fermion mass is not negligible,
these expressions differ only slightly \cite{APR1,BS}.

Of course, the above results are nothing new for QED, where they were
derived already a long time ago using the kinetic theory \cite{Sil}.
In that case, $\omega_0$ is
the {\it plasmon frequency}, at which the
electric charges oscillate in the plasma around their average positions.
Because of this analogy, it is possible to call $\omega_0$ the plasmon
frequency for the general $SU(N)$ gauge theory case. In fact, it is quite
remarkable that expressions (4.4) are so general.

In arbitrary covariant gauge with parameter $\xi$, the resummed gauge
propagator is (in the ITF)
\be
-i{\bf D}_{\mu\nu}(K) = {P_{\mu\nu} \over K^2 - \Pi_T(K)}
                      + {Q_{\mu\nu} \over K^2 - \Pi_L(K)}
                      - (\xi-1) {K_\mu K\nu \over K^4}
.\ee
The two poles of this resummed propagator define two propagation modes
(plasmons), which, because of the complicated structure of eqs.~(4.4), are
described by non-trivial dispersion relations. They are plotted in fig. ~6.
Notice that when $k\to 0$, they both tend to the same value,
$\omega=\omega_0$,
which can then be regarded as a {\it plasmon mass}.

There are also contributions below the light-cone, as the HTLs develop an
imaginary part there. In the static limit, one has
\bea
\Re\Pi_T(\omega\to 0,k) &=& k_D^2 {\omega^2\over k^2} ,\nonumber\\
\Re\Pi_L(\omega\to 0,k) &=& k_D^2,
\ena
where $k_D$ is the {\it Debye mass} corresponding to the screening of static
electric fields (again following the same analogy with QED). Indeed, one
finds that in non-degenerate plasmas $k_D^2=e^2 N_e/T$ and in degenerate
plasmas $k_D^2=(e^2/\pi^2)\mu p_F$. For ultra-relativistic plasmas,
$k_D^2=3\omega_0^2$. Notice that $k_D$ also appears in the expression for the
transverse propagator. However, it vanishes when $\omega\to 0$.
Hence, {\it there is no screening for space-like transverse gauge bosons}.
This has a very important consequence as the resummation of Braaten and
Pisarski is essentially based upon the presence of a mass term of order
$g^2T^2$ in the 2-point HTL. The transverse (or better called magnetic) mass
can only be of higher order, that is $m_{mag} = O(g^2 T)$. In fact, it is
exactly zero at all orders for QED \cite{GPY}. This is at the origin of the
infrared problem in hot gauge theories \cite{GPY,Lin2}. For an $SU(N)$ gauge
theory ($N>1$) and beyond some order in the perturbative
coupling constant, an infinite number of graphs contribute to the same order,
due to diagrams with multiple interacting space-like transverse gauge bosons.
In QED, or for a gauge theory at finite density, there is always enough
screening coming from the fermionic loops so the problem does not seem to
be so worrying. However, we shall see in the last section that there
might be an alternative to this problem.

Quite unfortunately, the literature contains may examples of confusion
between the Debye and the plasmon masses. They are totally unrelated.
The former acts as a screening mass for space-like longitudinal photons
(for instance in a fermion scattering on a target). The latter damps the
boson density in the plasma by a factor $e^{-\omega_0/T}$.

\subsection{The fermionic hard thermal self-energy}
Following the same procedure as in the last section, the fermionic HTL is
found to be
\be
\Sigma(K) = m_\beta^2 \( {1\over 2k}\ln{\omega+k\over \omega-k}\gamma_0
+ \( 1 - {\omega\over 2k}\ln{\omega+k\over \omega-k} \)
   {{\bf k.\gamma} \over k^2} \)
,\ee
where $m_\beta$ is another thermal mass given by
\be
m_\beta^2 = c_F {g^2 T^2\over 8}
,\ee
and $c_F=(N^2-1)/(2N)$ is a color factor.
In QED, one has to substitute $c_F g^2 \to e^2$.
By looking at the pole of the resummed propagator,
\be
-i{\bf S_F}(K) = {1\over \Ks -\Sigma(K)}
,\ee
one finds, unexpectedly, {\it two solutions}. They are plotted in fig.~7.
The first solution ($+$) corresponds obviously to the modified fermion
propagation in the plasma, but there does not seem to be any intuitive
physical interpretation for the second solution ($-$), apart from the fact
that it arrives as a pure collective effect \cite{Wel3}. This is a
completely new feature, which was not seen before the TFT was used \cite{BBS}.
Unfortunately, there seem to be no important phenomenological
consequences of this new mode, sometimes called ``plasmino''
\cite{Bra2}.

\subsection{Damping rates}
The resummation program of Braaten and Pisarski was in fact developed to
solve the long-standing problem of getting a gauge-independent answer for
the gluon damping rate \cite{BP3}.

The damping rates are computed through the discontinuity of
self-energies at the pole of the propagator and have the following physical
role: consider a particle distribution that is slightly out of
equilibrium. One has \cite{Wel4}
\be
{dn(\omega,t)\over dt} = -n(\omega,t) \Gamma_a(\omega)
                      + (1+\sigma n(\omega,t))\Gamma_c(\omega)
,\ee
where $\Gamma_a$ and $\Gamma_e$ are the absorption and creation rates
of the given particle, respectively. Notice the different statistical
factors for incoming and outgoing particles, with the parameter $\sigma$
to distinguish bosons ($\sigma=1$) from fermions ($\sigma=-1$).

Equation (4.11) has for general solution
\be
n(\omega,t) = {\Gamma_c \over \Gamma_a - \sigma\Gamma_c}
              +C(\omega) e^{-(\Gamma_a-\sigma\Gamma_c)t}
,\ee
where $C(\omega)$ is an arbitrary function, which does not depend on time.
Creation and absorption rates are related by the relation
\be
  \Gamma_a(\omega) = e^{\beta\omega} \Gamma_c(\omega)
,\ee
which is nothing but the KMS relation, written in a different form.
Using (4.13) we have
\be
n(\omega,t) = {1 \over e^{\beta\omega} - \sigma}
              + C(\omega) e^{-2\gamma t}
,\ee
where $\gamma= (\Gamma_a-\sigma\Gamma_c)/2$ is defined as the
{\it damping rate}. Therefore, its physical interpretation is rather clear.
It represents the inverse time scale it takes for a thermal distribution
to reach
equilibrium. In particular, the sign of the damping rate is of crucial
importance as, if negative, the system is obviously unstable.

One can therefore imagine the controversy, which has lasted for many years,
as the sign of the gluon damping rate was found to be dependent on the gauge
choice! Braaten and Pisarski have shown that in order to get a
consistent
result, one has to use their effective perturbation theory \cite{BP3}.

How to compute damping rates? The creation and absorption rates are
related to the imaginary part of the self-energy \cite{Wel4}. In the RTF,
Kobes and Semenoff have extended the cutting rules (or Cutkosky rules)
to the finite temperature case \cite{KS}. Again, results are similar in
the ITF and in the RTF.

{}From the results shown above, eqs.~(4.4) and (4.8), it is easy to see that
there is no HTL for the damping rates. Indeed, the HTLs have an imaginary
part only below the light-cone, whereas the pole of the propagator is
clearly above it. Therefore, the damping rates must be of
higher order, $g^2 T$, which makes them difficult to compute. Indeed, in
order to go beyond the leading order in the BPR, one has to subdivide
diagrams into different pieces, according to whether the internal lines
carry hard
or soft momenta. For soft external gluons, one has to use effective
vertices and propagators (see fig.~8). For transverse and longitudinal
gluons at rest (!) the result is \cite{BP3}
\be
  \gamma_T =\gamma_L \simeq 0.264013 g^2 T
.\ee
So the sign is positive and independent of the gauge, which implies that the
quark-gluon plasma is perturbatively stable.

The case of a hard external gluon turns out to be very interesting, as
there appears a singular behaviour that can only be cut by introducing a
higher order cut-off of the type $m_{mag}$ or $\gamma_T$ itself \cite{Pis3}.
In this case, effective vertices are not used and only one soft
internal line
is resummed. Only the transverse gluon damping rate diverges and is found to be
\cite{Pis3}
\be
 \gamma_T = {g^2 NT\over 8\pi} \( \ln{\omega_0^2\over m_{mag}^2 +
                                   2m_{mag}\gamma_T} + 1.09681\ldots \)
.\ee
Hence, $\gamma_T$ appears both on the left-hand side and on the right-hand
side of the equation. In order to get this result, one has to use a resummed
propagator also for the one hard internal line, which has nothing to do with
the
resummed propagator of Braaten and Pisarski. In fact, this propagator
does not resum the HTL, but some other contribution. The result is
also sensitive to the magnetic mass. This is bad in the sense that nobody
knows how to calculate this parameter consistently, except by going on the
lattice. However, this uncertainty shows up only at the logarithmic level
and has therefore very little phenomenological implications. The coefficient
in front of the logarithm is on the other hand perfectly well known, and is
consistently obtained by using the BPR.

The damping rates for fermions have similar features. What of hot QED,
where the magnetic mass vanishes? There, it seems that there is no
consistent way to get the damping rate \cite{Pis3,BN} and the problem is
still open. Among the possible different solutions, the simplest could be
that there is no pole at all in the fermion propagator \cite{PPS}!
 It is also
legitimate to ask the physical relevance of a quantity that is defined at
a complex energy! On the other hand, one may note
that the creation and absorption rates of fermions travelling through a heat
bath are perfectly defined  and are of order $g^2T\ln(1/g)$, as expected
\cite{APR2}.

\subsection{Final remarks}
When computing at finite temperature, there appear singularities as in
any other quantum field theory. Ultra-violet singularities are the same
as in vacuum and are disposed of by renormalization. Infrared
singularities occur when a momentum is vanishing and collinear
divergences when a line goes on shell. According to the
previous discussion, the singularity can be screened by using a resummed
propagator or vertex. However, terms that generate infrared singularities
are not the same as those that generate the HTLs.
Indeed, a typical on-shell expansion of the self-energy is
(at first order in $g$)
\be
\Sigma(K) = a g^2 T^2 \int dx {x \over e^x - 1}
          + b g^2 K^2 \int {dx\over x} {1\over e^x - 1}
,\ee
where $a$ and $b$ are numerical factors. The first term is proportional
to $T^2$ and represents the HTL. The second term is infrared-singular and
{\it a priori} infinite if one does not regularize in some way.
However, by resummation, we know: i) that there are thermal cut-offs of
order $gT$ and ii) that the particle on-shellness is restricted to
$K^2 \sim g^2 T^2$ whether it is soft or hard. Therefore, the second
contribution is of order $g^2\times g^2 T^2 (1/g) \sim g^3 T^2$ and
is smaller than the HTL. It is clear then, for the BPR to work, that
the term with the screened singularity must contribute less than the term
of the HTL. This happens to be true indeed, except of course in the case
where there is no screening, i.e., when the soft line is a magnetic
boson. On the other hand, it has also been found that those kinds of infrared
singularities can cancel out in the physical answer. Indeed, although a
complete general proof is still lacking, there is by now enough accumulated
evidence that the KLN theorem works also at finite temperature
\cite{Dilept1,AR,KLNT}.

Hence, the resummation of Braaten and Pisarski seems really to work well
and certainly oversteps its original goals. Even in the case when the weak
coupling limit is not justified, this separation between soft and hard
scales can be used as a mathematical trick, which is certainly more elegant
than any other method. I only want for proof the
numerous computations using these techniques [32-36,42,43,58,65,69].

\sect{Conclusion}
The list of references shows by its length that the TFT is now a mature
field,
with a growing activity over the last few years. A number of important
questions have already been answered, such as for instance
the comparisons between the
different formalisms. The main difficulties of the TFTs are of course infrared
singularities. A lot of effort has been done and the situation can be
grossly summarized as follows: one can make reliable perturbative
calculations when probing the thermal system at a scale $T$. By using
the resummation of Braaten and Pisarski one can go down to the scale $gT$.
The next step is to probe the magnetic mass scale, $g^2T$, onto which very
little is known.

There are many issues which are still unresolved. For instance, nobody knows
exactly what the behavior of the running coupling constant is at finite
temperature, and even if the question is relevant or not \cite{coupling}.

\bigskip
Finally, I could not end this paper without mentioning the interesting
connection between thermal and topological field theories as, for
instance, the fact that the Minkowski vacuum agrees with a thermal state
for an accelerated observer \cite{Geo}.

\bigskip\noindent
{\large\large Acknowledgements}

This paper is based upon lectures given at the IST (Lisbon, Nov. 92) and at
CERN (Geneva, Jan. 93). I thank both organisers, J.~Seixas and
L.~Alvarez-Gaum\'e, for the opportunity to discuss these issues with an
enthusiastic audience.

\newpage
{\large\bf Figure Captions\\[0.5cm]}
\begin{description}
\item[Fig.~1] The real-time contour in the complex-time plane.
\item[Fig.~2] The deformed contour for calculating (3.2) in the ITF.
\item[Fig.~3] The two topologies for the self-energy at 2-loop order.
\item[Fig.~4] A ``daisy'' diagram with an arbitrary number of loops
attached to the tadpole.
\item[Fig.~5] The effective propagators and vertices.
\item[Fig.~6] The dispersion relations for the bosonic modes.
\item[Fig.~7] The dispersion relations for the fermionic modes.
\item[Fig.~8] The resummed diagrams for the soft gluon damping rates.
\end{description}

\newpage
\begin{thebibliography}{99}
\bibitem{LvW} N. P. Landsman and Ch. G. van Weert, Phys. Rep. {\bf 145}
             (1987) 141.

\bibitem{Kap} J.~Kapusta, {\it Finite Temperature Field Theory} (Cambridge
             University Press, 1989).

\bibitem{Tol} R. C. Tolman, {\it The Principles of Statistical Mechanics}
              (Oxford University Press, 1938).

\bibitem{Kel} L. V. Keldysh, Sov. Phys. {\bf 20} (1964) 1018.

\bibitem{AUM} T. Arimitsu and H. Umezawa, Prog. Theor. Phys. {\bf 74} (1985)
              429, {\bf 77} (1987) 32 and {\bf 77} (1987) 53;
              H. Matsumoto, Prog. Theor. Phys. {\bf 79} (1988) 373.

\bibitem{Nie} A. J. Niemi, Phys. Lett. {\bf B203} (1988) 425.

\bibitem{CH} E. Calzetta and B. L. Hu, Phys. Rev. {\bf D37} (1988) 2878.

\bibitem{Wel1} H. A. Weldon, Ann. Phys. (NY) {\bf 193} (1989) 166 and
                                             {\bf 193} (1989) 177;
               A.~Ni\'egawa and A.~Takashiba, Proc. of the Workshop on High
               Temperature Field Theory, Winnipeg, July 1992;
        M.~Chaichian and I.~Sendo, Nucl. Phys. {\bf B396} (1993) 737.

\bibitem{Ich} S.~Ichimaru, {\it Plasma Physics, An Introduction to Statistical
            Physics of Charged Particles} (Benjamin-Cummings, Menlo Park,
1986).

\bibitem{LP} E. M. Lifshitz and L. P. Pitaevskii, {\it Statistical Physics}
             (Pergamon Press, Oxford, 1980).

\bibitem{FW} A. L. Fetter and J. D. Walecka, {\it Quantum Theory of
             Many-Particle Systems}, (McGraw-Hill, New York, 1971).

\bibitem{LeB} For a recent review, see M. Le Bellac, Lectures given at
the 30th Schladming Winter School, Schladming, Austria, 1991.

\bibitem{TFT1} Proc. of the Workshops on {\it Thermal Field Theories and
               their applications}:, Physica {\bf A158} (1989) 1-558 and
               H.~Ezawa, T.~Arimitsu and Y.~Mashimoto, Elsevier, Amsterdam,
               1991.

\bibitem{Sch} J. Schwinger, J. Math. Phys. {\bf 2} (1961) 407.

\bibitem{Mil} R. Mills, {\it Propagators for Many-Particle Systems} (Gordon and
              Breach, New York, 1969).

\bibitem{KSW} R. L. Kobes, G. W. Semenoff and N. Weiss, Z. Phys. {\bf C29}
              (1985) 371.

\bibitem{Haa} R. Haag, Commun. Math. Phys. {\bf 5} (1967) 215.

\bibitem{Eza} H. Ezawa, in {\it Quantum Field Theory}, ed. F. Mancini
              (North-Holland, Amsterdam, 1986).

\bibitem{Mat} T. Matsubara, Prog. Theor. Phys. {\bf 14} (1955) 351.

\bibitem{Ume} H. Umezawa, H. Matsumoto and M. Tachiki, {\it Thermo Field
              Dynamics and Condensed States} (North Holland, Amsterdam, 1982).

\bibitem{Oji} I. Ojima, Ann. Phys. (NY) {\bf 137} (1981) 1.

\bibitem{EE} K.~Enqvist and K.~Eskola, Mod. Phys. Lett. {\bf A5} (1990) 1919.

\bibitem{COBE} J.~C.~Mather et al., Astrophy. J. {\bf 354} (1990) L37;
               G.~F.~Smoot et al., Astrophys. J. {\bf 396} (1992) L1.

\bibitem{Wei} S. Weinberg, {\it Cosmology and Gravitation} (McGraw-Hill,
              New York, 1970).

\bibitem{Ber} C. W. Bernard, Phys. Rev. {\bf D12} (1974) 3312.

\bibitem{DJ} L. Dolan and R. Jackiw, Phys. Rev. {\bf D9} (1974) 3320.

\bibitem{Lin1} A.~Linde, Rep. Prog. Phys. {\bf 42} (1979) 389.

\bibitem{Oli} K. Olive, Phys. Rep. {\bf 190} (1990) 307.

\bibitem{Lei} For a list of references, see R. G. Leigh, Proceedings of the
26th International Conference  on High Energy Physics, Dallas, 1992.

\bibitem{BPPS} R. Baier, E. Pilon, B. Pire and D. Schiff, Nucl. Phys.
               {\bf B336} (1990) 157.

\bibitem{Reb} A.~Rebhan, Nucl. Phys. {\bf B351} (1991) 706 and {\bf B368}
              (1992) 479; U.~Kraemmer and A.~Rebhan, Phys. Rev. Lett.
              {\bf 67} (1991) 793.

\bibitem{AK} T.~Altherr and K.~Kainulainen, Phys. Lett. {\bf B262} (1991) 79.

\bibitem{Alt1}  T.~Altherr, Z. Phys. {\bf C47} (1990) 559 and Ann. Phys. (NY)
             {\bf 207} (1991) 374.

\bibitem{BY} E.~Braaten and T.~C.~Yuan, Phys. Rev. Lett. {\bf 66} (1991) 2183.

\bibitem{Bra1} E.~Braaten, Phys. Rev. Lett. {\bf 66} (1991) 1655.

\bibitem{AKr} T. Altherr and U. Kraemmer, Astroparticle Phys. {\bf 1}
              (1992) 133.

\bibitem{APR1} T.~Altherr, E.~Petitgirard and T.~del Rio Gaztelurrutia,
               to appear in Astroparticle Physics.

\bibitem{BS} E.~Braaten and D.~Segel, Northwestern Univ. preprint
             NUHEP-TH-93-1.

\bibitem{CGS} J.~Cleymans, R.~V.~Gavai and E.~Suhonen, Phys. Rep. {\bf 130}
            (1986) 217.

\bibitem{Sat} H. Satz, Annu. Rev. Nucl. Part. Sci. {\bf 35} (1985) 245.

\bibitem{Dilept1} R.~Baier, B.~Pire and D.~Schiff, Phys. Rev. {\bf D38}
        (1988) 2814;
        T.~Altherr, P.~Aurenche and T.~Becherrawy, Nucl. Phys. {\bf B315}
        (1989) 436;
        T.~Altherr and P.~Aurenche, Z. Phys. {\bf C45} (1989) 99;
        T.~Altherr and T.~Becherrawy, Nucl. Phys. {\bf B330} (1990) 174;
        Y.~Gabellini, T.~Grandou and D.~Poizat, Ann. Phys. (NY)
        {\bf 202} (1990) 436.

\bibitem{AR} T.~Altherr and P.~V.~Ruuskanen,
        Nucl. Phys. {\bf B380} (1992) 377.

\bibitem{Dilept2} E.~Braaten, R.~D.~Pisarski and T.~Chiang~Yuan,
        Phys. Rev. Lett. {\bf 64} (1990) 2242;
        E.~Braaten and M.~Thoma, Phys. Rev. {\bf D44} (1991) 1298;
        J.~Kapusta, P.~Lichard and D.~Seibert, Phys. Rev. {\bf D44} (1991)
2744;
        R.~Baier, H.~Nakkagawa, A.~Ni\'egawa and K.~Redlich,
        Z. Phys. {\bf C53} (1992) 433.

\bibitem{Pis1} R.~D.~Pisarski, Nucl. Phys. {\bf B309} (1988) 476.

\bibitem{Eva1} T.~S.~Evans, Phys. Rev. {\bf D47} (1993) R4196.

\bibitem{Abr} A.~A.~Abrikosov, Mod. Phys. Lett. {\bf A5} (1990) 2183.

\bibitem{AALBP} T.~Altherr and P.~Aurenche, Phys. Rev. {\bf D40} (1989) 4171;
                M.~Le~Bellac and D.~Poizat, Z. Phys. {\bf C47} (1990) 125.

\bibitem{Lan} N.~P.~Landsmann, Ann. Phys. (NY) {\bf 186} (1988) 141.

\bibitem{FGN} Y.~Fujimoto, R.~Grigjanis and H.~Nishino, Phys. Lett.
              {\bf B141} (1984) 83;
              Y. Fujimoto and R. Grigjanis, Z. Phys. {\bf C28} (1985) 395.

\bibitem{RTFITF} R.~Kobes, Phys. Rev. {\bf D42} (1990) 562,
                 Phys. Rev. Lett. {\bf 67} (1991) 1384;
                 T.~S.~Evans, Phys. Lett. {\bf B249} (1990) 286 and
                 {\bf B252} (1990) 108, and Nucl. Phys. {\bf B374} (1992) 340;
                 P.~Aurenche and T.~Becherrawy, Nucl. Phys. {\bf B379}
                 (1992) 259;
                 M.~A.~van Eijck and Ch.~G.~van Weert, Phys. Lett.
                 {\bf B278} (1992) 305.

\bibitem{Alt2} T.~Altherr, Phys. Lett. {\bf B238} (1990) 360.

\bibitem{KKR} R.~Kobes, G.~Kunstater and A.~Rebhan, Phys. Rev. Lett.
             {\bf 64} (1990) 2992 and Nucl. Phys. {\bf B355} (1991) 1.

\bibitem{Kin} T.~Kinoshita, J. Math. Phys. {\bf 3} (1962) 650.

\bibitem{LN} T.~D.~Lee and M.~Nauenberg, Phys. Rev. {\bf 133} (1964) 1549.

\bibitem{AGP} T.~Altherr, T.~Grandou and R.~D.~Pisarski, Phys. Lett.
             {\bf B271} (1991) 183.

\bibitem{Pis2} R.~D.~Pisarski, Phys. Rev. Lett. {\bf 63} (1989) 1129.

\bibitem{BP1} E.~Braaten and R.~D.~Pisarski, Nucl. Phys. {\bf B337} (1990) 569
             and {\bf B339} (1990) 310.

\bibitem{BP2} E.~Braaten and R.~D.~Pisarski, Phys. Rev. {\bf D45} (1992)
1827.

\bibitem{Wel2} H.~A.~Weldon, Phys. Rev. {\bf D26} (1982) 1394.

\bibitem{Sil} V.~P.~Silin, Sov. Phys. JETP {\bf 11} (1960) 1136.

\bibitem{GPY} D.~J.~Gross, R.~D.~Pisarski and L.~G.~Yaffe, Rev. Mod. Phys.
             {\bf 53} (1981) 43.

\bibitem{Lin2} A.~Linde, Phys. Lett. {\bf B93} (1980) 327.

\bibitem{Wel3} H.~A.~Weldon, Physica {\bf A158} (1989) 169.

\bibitem{BBS} G.~Baym, J.-P.~Blaizot and B. Svetitsky, Phys. Rev. {\bf D46}
              (1992) 4043.

\bibitem{Bra2} E.~Braaten, Astrophys. J. {\bf 392} (1992) 70.

\bibitem{BP3} E.~Braaten and R.~D.~Pisarski, Phys. Rev. {\bf D42} (1990)
2156.

\bibitem{Wel4} H.~A.~Weldon, Phys. Rev. {\bf D28} (1983) 2007.

\bibitem{KS} R.~L.~Kobes and G.~W.~Semenoff, Nucl. Phys. {\bf B260} (1985) 714
             and {\bf B272} (1986) 329;
             N.~Ashida, H.~Nakkagawa, A.~Ni\'egawa and H.~Yokota, Ann.
             Phys. (NY) {\bf 215} (1992) 315.

\bibitem{Pis3} R.~D.~Pisarski, Phys. Rev. {\bf D47} (1993) 5859.

\bibitem{BN} R.~Baier and A.~Ni\'egawa, Proc. of the Workshop on High
Temperature Field Theory, Winnipeg, July 1992.

\bibitem{PPS} S.~Peign\'e, E.~Pilon and D.~Schiff, Orsay Univ. preprint
              LPTHE 93-13.

\bibitem{APR2} T.~Altherr, E.~Petitgirard and T.~del Rio Gaztelurrutia,
               Phys. Rev. {\bf D47} (1993) 703.

\bibitem{KLNT} T.~Grandou, M.~Le~Bellac and J.~L.~Meunier, Z. Phys.
              {\bf C45} (1989) 575;
               T.~Grandou, M.~Le~Bellac and D.~Poizat, Phys. Lett.
              {\bf B249} (1990) 478 and Nucl. Phys. {\bf B358} (1991) 408;
               T.~Altherr, Phys. Lett. {\bf B262} (1991) 314;
               H.~A.~Weldon, Phys. Rev. {\bf D44} (1991) 3955;
               M.~Le~Bellac and P.~Reynaud, Nucl. Phys. {\bf B380} (1992)
               423;
               A. Ni\'egawa and K. Takashiba, Nucl. Phys. {\bf B370} (1992)
335;
               T.~Altherr and T.~Grandou, to appear in Nucl. Phys. B;
               A.~Ni\'egawa, Osaka Univ. preprint, OCU-PHYS-149, May 1993.

\bibitem{coupling} K.~Enqvist and K.~Kainulainen, Z. Phys. {\bf C53} (1992) 87;
                   P.~Aurenche, E.~Petitgirard and T.~del Rio
                   Gaztelurrutia, Phys. Lett. {\bf B297} (1993) 337.

\bibitem{Geo} 
              Y.-X.~Gui, Phys. Rev. {\bf D46} (1992) 46.
\end{thebibliography}
\end{document}